\documentclass[nofootinbib,10pt,twocolumn,pra,showpacs,tightenlines, showkeys, english,aps]{revtex4}
\usepackage[T1]{fontenc}
\usepackage[latin9]{inputenc}
\usepackage{color}
\usepackage{float}
\usepackage{epsfig,graphicx,times}
\usepackage{amstext}
\usepackage{amsmath}  
\usepackage{amssymb}
\usepackage{stackrel}
\usepackage{graphicx}
\usepackage{esint}
\usepackage{epstopdf}
\usepackage{epsfig}

\makeatletter
\@ifundefined{textcolor}{}
{%
 \definecolor{BLACK}{gray}{0}
 \definecolor{WHITE}{gray}{1}
 \definecolor{RED}{rgb}{1,0,0}
 \definecolor{GREEN}{rgb}{0,1,0}
 \definecolor{BLUE}{rgb}{0,0,1}
 \definecolor{CYAN}{cmyk}{1,0,0,0}
 \definecolor{MAGENTA}{cmyk}{0,1,0,0}
 \definecolor{YELLOW}{cmyk}{0,0,1,0}
}

\makeatother

\usepackage{babel}
\begin{document}

\title{Higher order  two-mode and multi-mode entanglement in Raman processes }

\author{Sandip Kumar Giri$^{1,2}$, Biswajit Sen$^{3}$ , Anirban Pathak$^{4}$,
Paresh Chandra Jana$^{2}$}

\affiliation{$^{1}$Department of Physics, Panskura Banamali College, Panskura-721152,
India\\
$^{2}$Department of Physics, Vidyasagar University, Midnapore-721102,
India\\
$^{3}$Department of Physics, Vidyasagar Teachers' Training College,
Midnapore-721101, India\\
$^{4}$Jaypee Institute of Information Technology, A-10, Sector-62,
Noida, UP-201307, India}
\begin{abstract}
The existence of higher order entanglement in the stimulated and spontaneous
Raman processes is established using the perturbative solutions of
the Heisenberg equations of motion for various field modes that are
obtained using the Sen-Mandal technique and a fully quantum mechanical
Hamiltonian that describes the stimulated and spontaneous Raman processes.
Specifically, the perturbative Sen-Mandal solutions are exploited
here to show the signature of the higher order two-mode and multi-mode
entanglement. In some special cases, we have also observed higher
order entanglement in the partially spontaneous Raman processes. Further,
it is shown that the depth of the nonclassicality indicators (parameters)
can be manipulated by the specific choice of coupling constants, and
it is observed that the depth of nonclassicality parameters increases with the
order. 
\end{abstract}



\maketitle

\section{Introduction}

\begin{figure}[h]
\centering{} \includegraphics[angle=-90,scale=0.9]{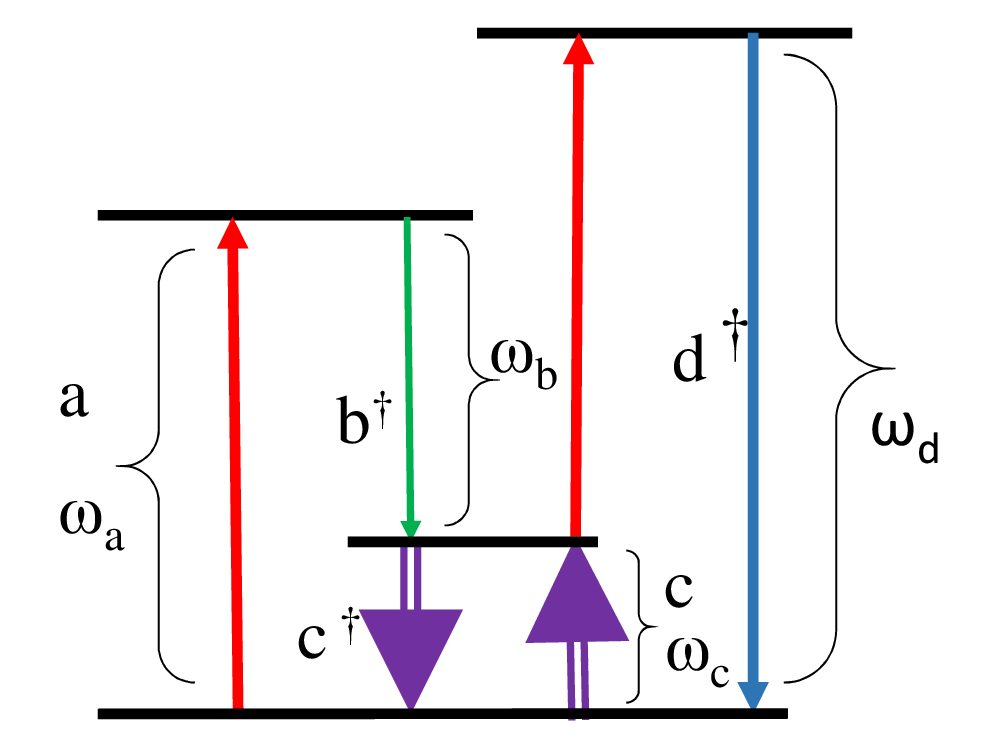} \caption{\label{fig:scheme}(Color online) Two-photon stimulated Raman scheme.
The pump photon is converted into a Stokes photon and a phonon. The
pump photon can also mix with a phonon to produce an anti-Stokes photon.}
\end{figure}

With the advent of quantum computation and communication, entanglement
has appeared as a very important resource \cite{Bennet1993,densecoding,Shukla,my-book}.
For example, its essential role in many processes, such as teleportation
\cite{Bennet1993}, dense coding \cite{densecoding}, quantum information
splitting \cite{Shukla}, etc., are now well established. In short,
entangled states are required to perform various important tasks related
to quantum information processing. Entanglement is produced in many
physical systems and there exists a large number of criteria for detection
of entanglement (\cite{phys rep rev} and references therein). The
first inseparability criterion was proposed by Peres \cite{peres}
in 1996. Since then several inseparability inequalities have been
reported for two mode and multi-mode states \cite{duan,hungh,lee,simon,hz-prl,two-mode-citeria-hz,hill-dung-zheng-pra-muti-partite,GSA-Ashoka,nonclassical corr,two-photon laser,A-Miranowicz-et,application,Adam1}.
For the present study, we have mostly used higher order version of
two criteria of Hillery and Zubairy \cite{hz-prl,two-mode-citeria-hz}.
To be precise, we have used these criteria to investigate the existence
of higher order entanglement in Raman processes, as depicted in Fig.
\ref{fig:scheme}. From Fig. \ref{fig:scheme} we can easily observe
that the scheme illustrated here is essentially a sequential double
Raman process \cite{nonclassical corr}. Nonclassical properties of
this system are studied since long (for a review see Ref. \cite{Adam2}). Initial studies on this system
were restricted to the short-time approximation \cite{szlachetka1,perina}.
However, recently nonclassical properties of this system have been
investigated by some of us \cite{Anirban with Perina,raman-pra} using
different approaches other than short-time approximation, but the
possibility of observing higher order entanglement is not investigated
in any of the existing studies. Further, several applications of Raman
processes have been reported in the recent past \cite{quanreap1,quanreap2}
and higher order nonclassicality in different physical systems have
also been reported experimentally \cite{bondani1,bondani2,chekhova,ent-in-morethan40atoms-science}
and theoretically \cite{sandip-bectwomode,kishore-cocoupler,sandip-atommolecule,kishore-contra}.
Keeping these facts in mind, present paper aims to investigate the
possibility of higher order entanglement in the spontaneous, partially
spontaneous and stimulated Raman processes and effect of the phase
of the pump mode on the higher order entanglement. In what follows,
Raman process is described as shown in Fig. \ref{fig:scheme} and
a completely quantum mechanical description of the system is used
to obtain analytic expressions for the time evolution of the various
filed modes involved in the process. The expressions are obtained
using a perturbative method known as Sen-Mandal method \cite{bsen1,bsen2,bsen3,bsen4}.
Subsequently, the expressions obtained using this method and Hillery-Zubairy
criteria \cite{hz-prl,two-mode-citeria-hz} are used to investigate
the existence of multi-mode entanglement and higher order two-mode
entanglement. Interestingly, the investigation has revealed the existence
of multi-mode entanglement (which is essentially higher order as is
witnessed via higher order correlation function) and higher order
two-mode entanglement involving various modes present in Raman process.

Remaining part of the present paper is organized as follows. In Section
\ref{sec:Model-Hamiltonian}, the Hamiltonian of stimulated Raman
processes and its operator solution are briefly described. In Section
\ref{sec:Intermodal-entanglement}, the solution is used to show the
existence of higher order two-mode, three-mode, and four-mode entanglement
and the effect of phase of the pump mode on the higher order entanglement.
Finally, the paper is concluded in Section \ref{sec:Conclusions}.

\section{Model Hamiltonian\label{sec:Model-Hamiltonian}}

A completely quantum mechanical description of stimulated and spontaneous
Raman processes described in Fig. \ref{fig:scheme} is given by the
Hamiltonian \cite{szlachetka1,szlachetka2,raman-pra,walls,bsen1,bsen2,bsen3,bsen4} 

\begin{equation}
\begin{array}{lcl}
H & = & \omega_{a}a^{\dagger}a+\omega_{b}b^{\dagger}b+\omega_{c}c^{\dagger}c+\omega_{d}d^{\dagger}d\\
 & + & g\left(ab^{\dagger}c^{\dagger}+h.c.\right)+\chi\left(acd^{\dagger}+\mathrm{h.c.}\right),
\end{array}\label{eq:hamiltonian}
\end{equation}
where h.c. stands for the Hermitian conjugate. Throughout the present
paper, we use $\hbar=1$. The annihilation (creation) operators $a(a^{\dagger}),\, b(b^{\dagger}),\, c\left(c^{\dagger}\right),\, d(d^{\dagger})$
correspond to the laser (pump) mode, Stokes mode, vibration (phonon)
mode and anti-Stokes mode, respectively. They obey the well-known
bosonic commutation relations. The frequencies $\omega_{a}$, $\omega_{b}$,
$\omega_{c}$ and $\omega_{d}$ correspond to the frequencies of pump
mode $a$, Stokes mode $b$, vibration (phonon) mode $c$ and anti-Stokes
mode $d$, respectively. The parameters $g$ and $\chi$ are the Stokes
and anti-Stokes coupling constants, respectively. Coupling constant
$g$ ($\chi$) denotes the strength of coupling between the Stokes
(anti-Stokes) mode and the vibrational (phonon) mode and depends on
the actual interaction mechanism. The dimension of $g$ and $\chi$
are that of frequency and consequently $gt$ and $\chi t$ are dimensionless.
Further, $gt$ and $\chi t$ are very small compared to unity. Further,
we would like to note that in our present study, only one vibration
(phonon) mode has been considered for the mathematical simplicity.
In order to study the possibility of the existence of higher order
entanglement, we need simultaneous solutions of the following Heisenberg
operator equations of motion for various field operators: 
\begin{equation}
\begin{array}{lcl}
\dot{a} & = & -i\left(\omega_{a}a+gbc+\chi cd\right)\\
\dot{b} &  = & -i\left(\omega_{b}b+gac^{\dagger}\right)\\
\dot{c} & = & -i\left(\omega_{c}c+gab^{\dagger}+\chi a^{\dagger}d\right)\\
\dot{d} &  = &  -i\left(\omega_{d}d+\chi ac\right).
\end{array}\label{eq:equn of motion}
\end{equation}
The above set of equations (\ref{eq:equn of motion}) is coupled nonlinear
differential equations of filed operator and are not exactly solvable
in the closed analytical form under weak pump condition. However,
for the very strong pump, the operator $a$ can be replaced by a $c$
number and these equations (\ref{eq:equn of motion}) are exactly
solvable in that case \cite{perina}. In order to solve these equations
under weak pump approximation, we have used Sen-Mandal perturbative
approach \cite{bsen1,bsen2,bsen3,bsen4}. The solutions obtained using
this approach are more general than the one obtained for the same
system using well-known short-time approximation. Details of the calculations
are given in our previous papers \cite{bsen1,bsen2,bsen3,bsen4}.
Here we just note that under weak pump approximation, the solutions
of Eq. (\ref{eq:equn of motion}) assume the following form: \begin{widetext}
\begin{equation}
\begin{array}{lcl}
a(t) & = & f_{1}a(0)+f_{2}b(0)c(0)+f_{3}c^{\dagger}(0)d(0)+f_{4}a^{\dagger}(0)b(0)d(0)+f_{5}a(0)b(0)b^{\dagger}(0)\\
 & + & f_{6}a(0)c^{\dagger}(0)c(0)+f_{7}a(0)c^{\dagger}(0)c(0)+f_{8}a(0)d^{\dagger}(0)d(0),\\
b(t) & = & g_{1}b(0)+g_{2}a(0)c^{\dagger}(0)+g_{3}a^{2}(0)d^{\dagger}(0)+g_{4}c^{\dagger^{2}}(0)d(0)+g_{5}b(0)c(0)c^{\dagger}(0)\\
 & + & g_{6}b(0)a(0)a^{\dagger}(0),\\
c(t) & = & h_{1}c(0)+h_{2}a(0)b^{\dagger}(0)+h_{3}a^{\dagger}(0)d(0)+h_{4}b^{\dagger}(0)c^{\dagger}(0)d(0)+h_{5}c(0)a(0)a^{\dagger}(0)\\
 & + & h_{6}c(0)b(0)b^{\dagger}(0)+h_{7}c(0)d^{\dagger}(0)d(0)+h_{8}c(0)a^{\dagger}(0)a(0),\\
d(t) & = & l_{1}d(0)+l_{2}a(0)c(0)+l_{3}a^{2}(0)b^{\dagger}(0)+l_{4}b(0)c^{2}(0)+l_{5}c^{\dagger}(0)c(0)d(0)\\
 & + & l_{6}a(0)a^{\dagger}(0)d(0).
\end{array}.\label{soln1}
\end{equation}
 The parameters $f_{i},\, g_{i},\, h_{i}$ and $l_{i}$ are computed
the initial boundary conditions. In order to obtain the solutions
we use the boundary condition as at $t=0$, in the first term of the
Eq. (\ref{soln1}). It is clear that $f_{1}(0)=g_{1}(0)=h_{1}(0)=l_{1}(0)=1$
and $f_{i}(0)=g_{i}(0)=h_{i}(0)=l_{i}(0)=0$ (for $i=2,\,3,\,4,\,5,\,6,\,7$
and $8$). Under these initial conditions the corresponding solutions
for $f_{i}(t),\, g_{i}(t),\, h_{i}(t)$ and $l_{i}(t)$ are already
reported in our earlier work \cite{bsen1,bsen2,bsen3,bsen4}. The
same is included here as Appendix \ref{appendix}. \end{widetext}

The solution (\ref{soln1}), is valid up to the second orders in $g$
and $\chi$. In what follows, we consider $\Delta\omega_{1}=\omega_{b}+\omega_{c}-\omega_{a}$
and $\Delta\omega_{2}=\omega_{a}+\omega_{c}-\omega_{d}$. The detunings
$\Delta\omega_{1}$ and $\Delta\omega_{2}$ are usually very small.
In the present work we have chosen $|\Delta\omega_{1}|=0.1$ MHz and
$|\Delta\omega_{2}|=0.19$ MHz.

\section{Higher order intermodal entanglement\label{sec:Intermodal-entanglement}}

In order to investigate the higher-order entanglement in spontaneous
and stimulated Raman processes, we assume that all photon and phonon
modes are initially coherent. In other words, the composite boson
field consisting of photons and phonon is in an initial state which
is product of coherent states. Therefore, the composite coherent state
arises from the product of the coherent states $|\alpha_{1}\rangle,\,|\alpha_{2}\rangle$,
$|\alpha_{3}\rangle,$ and $|\alpha_{4}\rangle$ which are the eigenkets
of $a,\, b,\, c$ and $d$ respectively. Thus, the initial composite
state is 
\begin{equation}
|\psi(0)\rangle=|\alpha_{1}\rangle\otimes|\alpha_{2}\rangle\otimes|\alpha_{3}\rangle\otimes|\alpha_{4}\rangle.\label{eq:initial state}
\end{equation}
It is clear that the initial state is separable. Now the field operator
$a(0)$ operating on such a composite coherent state gives rise to
the complex eigenvalue $\alpha_{1}.$ Hence we have, 
\begin{equation}
a(0)|\psi(0)\rangle=\alpha_{1}|\psi(0)\rangle,\label{3.7}
\end{equation}
where $|\alpha_{1}|^{2}$ is the number of input photons in the pump
mode. In a similar fashion, we can also describe three more complex
amplitudes $\alpha_{2}(t)$, $\alpha_{3}(t)$ and $\alpha_{4}(t)$
corresponding to the Stokes, vibrational (phonon) and anti-Stokes
field mode operators $b,$ $c$ and $d$, respectively. It is clear
that for a spontaneous process, the complex amplitudes except for
the pump mode, are necessarily zero. Thus, in the spontaneous Raman
process, $\alpha_{2}=\alpha_{3}=\alpha_{4}=0$ and $\alpha_{1}\neq0.$
For partially spontaneous process \cite{Anirban with Perina}, the
complex amplitude $\alpha_{1}$ and any one/two of the remaining three
eigenvalues are nonzero while the other two/one complex amplitudes
are/is zero. In the present investigation, we consider that the eigenvalue
corresponding to the pump mode is complex i.e., $\alpha_{1}=\left|\alpha_{1}\right|e^{-i\phi}$,
where $\phi$ is the phase angle, but the other eigenvalues (i.e.,
eigenvalues for the Stokes, vibrational (phonon) and anti-Stokes modes)
are real.

\subsection{Higher order two mode entanglement \label{sub:Two-mode-entanglement}}

In order to investigate the higher order two mode entanglement, we
use two criteria due to Hillery and Zubairy \cite{hz-prl,two-mode-citeria-hz}.
The first criteria of Hillery and Zubairy is 
\begin{equation}
E_{i,j}^{n,m}=\langle i^{\dagger n}i^{n}j^{\dagger m}j^{m}\rangle-|\langle i^{n}j^{\dagger m}\rangle|^{2}<0,\label{eq:HZ-1}
\end{equation}
and the second criterion is 
\begin{equation}
E_{i,j}^{\prime n,m}=\langle i^{\dagger n}i^{n}\rangle\langle j^{\dagger m}j^{m}\rangle-|\langle i^{n}j^{m}\rangle|^{2}<0.\label{eq:HZ-2}
\end{equation}
where$i$ and $j$ are any two arbitrary operators and $i,j\in\{a,b,c,d\}\forall i\neq j.$
Here $m$ and $n$ are the positive integers and the lowest possible
values of $m$ and $n$ are $m=n=1,$ which corresponds to the normal
(lowest) order intermodal entanglement. A quantum state is said to
be higher order entangled (bi-partite) if it is found to satisfy the
equation (\ref{eq:HZ-1}) and/or equation (\ref{eq:HZ-2}) for any
choice of the integers $m$ and $n$ satisfying $m+n\geq3.$ From
here onward we will refer to these criteria (\ref{eq:HZ-1}) and (\ref{eq:HZ-2})
as HZ-1 criterion and HZ-2 criterion, respectively. More specifically,
a higher order entangled state is one which is witnessed via a higher
order (order $k>2)$ correlation function and as per this definition
all multi-partite (multi-mode) entangled states are also higher order
entangled.

Before we proceed further, we note that these two criteria are only
sufficient (not necessary) for detection of entanglement. Keeping
this fact in mind, we have applied both of these two criteria to investigate
the existence of higher order intermodal entanglement between various
modes and have observed higher order intermodal entanglement in various
situations. In what follows, we have also investigated the possibility
of observing 3-mode and 4-mode entanglement.

\begin{widetext}

\begin{figure}[h]
\includegraphics[angle=-90,scale=0.7]{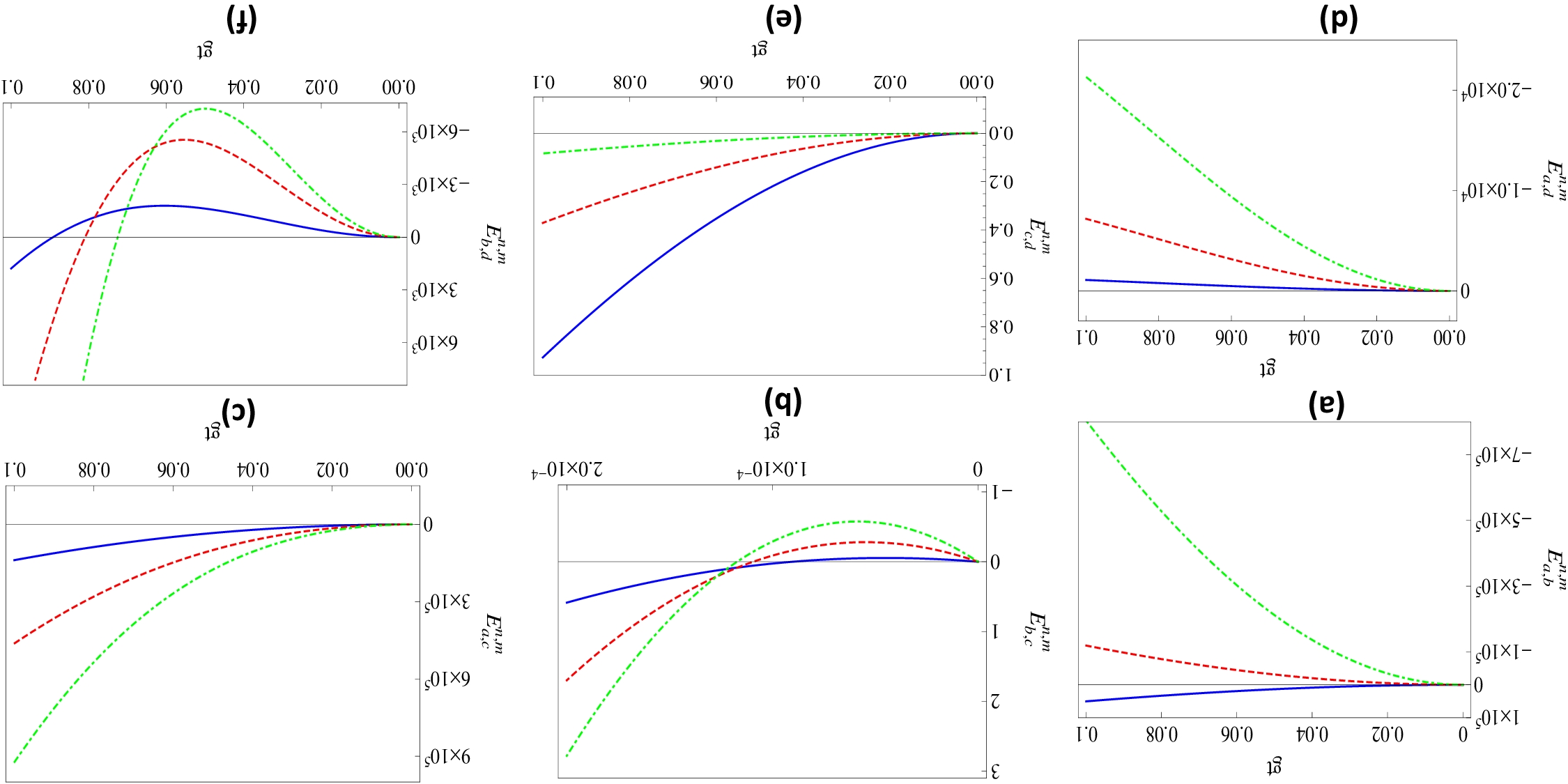}

\protect \caption{\label{fig:HZ2-1}(Colour online) Higher order intermodal entanglement
in stimulated Raman process with $\left|\alpha_{1}\right|=10,$ $\left|\alpha_{2}\right|=8,$
$\left|\alpha_{3}\right|=0.01$ and $\left|\alpha_{4}\right|=1$ using
HZ-1 criteron is shown for diffrent values of phase angle ($\phi=0,$
$\frac{\pi}{2}$ and $\pi$) in pump mode. Higher order intermodal
entanglement is observed in (a) pump-Stokes mode for phase angle $0,$
(b) Stokes-vibration phonon mode for phase angle $\phi=\frac{\pi}{2},$
(d) pump-anti-Stokes mode for phase angle $0,$ (f) Stokes-anti-Stokes
mode for phase angle $0$; and not observed in (c) pump-vibration
phonon mode and (e) vibration phonon-anti-Stokes mode. In all the
plots, the smooth line, dotted line and dash-dotted line are used
for the $m=1$ and $n=1,$ $2$ and $3$, respectively. In (e), $n=2$ and 3 are multiplied by $10^3$ and $10^6$, respectively. While in all the remaining cases, $n=1$ and 2 are shown 1500 and 50 times, respectively. }
\end{figure}

\end{widetext}

\begin{widetext}

\begin{figure}[h]
\includegraphics[angle=-90,scale=0.7]{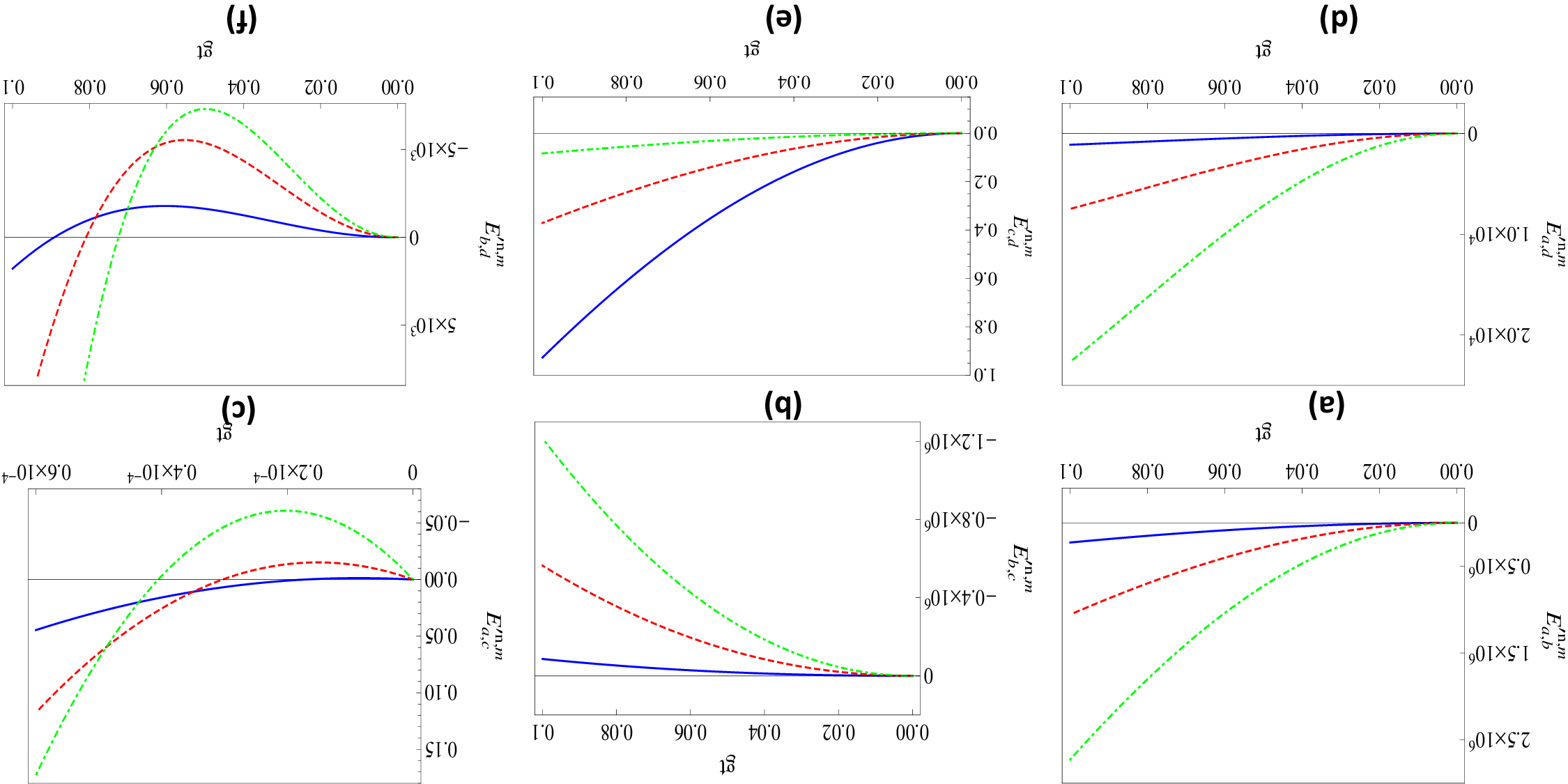}

\caption{\label{fig:HZ2} (Colour online) Higher order intermodal entanglement
in stimulated Raman process is illustrated using HZ-2 criterion with
different phase angle ($\phi=0,$ $\frac{\pi}{2}$ and $\pi$) in
pump mode for $\left|\alpha_{1}\right|=10,$ $\left|\alpha_{2}\right|=8,$
$\left|\alpha_{3}\right|=0.01$ and $\left|\alpha_{4}\right|=1$.
Specifically, higher order intermodal entanglement is observed in
(b) for Stokes-vibration phonon mode with phase angle $\phi=\frac{\pi}{2},$
(c) for pump-vibration phonon mode with phase angle $\phi=\frac{\pi}{2},$
and (f) for Stokes-anti-Stokes mode with phase angle $\frac{\pi}{2}$.
However, in (a), (d) and (e) higher order intermodal entanglement
is not observed for pump-Stokes, pump-anti-Stokes and vibration phonon-anti-Stokes
modes, respectively. The smooth, dotted and dash-dotted lines are
used for $m=1$ and $n=1,$ $2$ and $3$, respectively. Here, for (e) $n=2$ and 3 are multiplied by $10^3$ and $10^6$, respectively. For all the remaining cases, $n=1$ and 2 are shown 1500 and 50 times, respectively.}
\end{figure}

\end{widetext}

Let us first investigate the possibility of two mode entanglement
in Raman process using HZ-1 and HZ-2 criteria. From Eqs. (\ref{soln1}),
(\ref{eq:initial state}), (\ref{eq:HZ-1}) and (\ref{eq:HZ-2}),
we obtain the expression for the intermodal entanglement in pump and
Stokes mode as \begin{widetext}

\begin{equation}
\begin{array}{lcl}
\left(\begin{array}{c}
E_{a,b}^{n,m}\\
E_{a,b}^{\prime n,m}
\end{array}\right) & = & \left|f_{2}\right|^{2}m\left(m\left|\alpha_{1}\right|^{2\left(n+1\right)}\left|\alpha_{2}\right|^{2\left(m-1\right)}\mp n\left|\alpha_{1}\right|^{2n}\left|\alpha_{2}\right|^{2m}\right)+\left|f_{3}\right|^{2}n^{2}\left|\alpha_{1}\right|^{2\left(n-1\right)}\left|\alpha_{2}\right|^{2m}\left|\alpha_{4}\right|^{2}\end{array}.\label{ab}
\end{equation}
In the similar manner, for the remaining cases, we obtain expressions
for $E_{i,j}^{n,m}$ and $E_{i,j}^{\prime n,m}:i,j\in\{a,b,c,d\}\forall i\neq j$
using HZ-1 and HZ-2 criteria as follows

\begin{equation}
\begin{array}{lcl}
\left(\begin{array}{c}
E_{b,c}^{n,m}\\
E_{b,c}^{\prime nm,}
\end{array}\right) & = & \left|g_{2}\right|^{2}\left|\alpha_{2}\right|^{2\left(n-1\right)}\left|\alpha_{3}\right|^{2\left(m-1\right)}\left\{ n^{2}\left(1\pm2m\right)\left|\alpha_{1}\right|^{2}\left|\alpha_{3}\right|^{2}+m^{2}\left(1\pm2n\right)\left|\alpha_{1}\right|^{2}\left|\alpha_{2}\right|^{2}\right.\\
 & \pm & \left.m^{2}n^{2}\left|\alpha_{1}\right|^{2}\mp mn\left|\alpha_{2}\right|^{2}\left|\alpha_{3}\right|^{2}\right\} +\left|h_{3}\right|^{2}m^{2}\left|\alpha_{2}\right|^{2n}\left|\alpha_{3}\right|^{2\left(m-1\right)}\left|\alpha_{4}\right|^{2}\\
 & \pm & \left[g_{1}g_{2}^{\star}mn\left|\alpha_{2}\right|^{2\left(n-1\right)}\left|\alpha_{3}\right|^{2\left(m-1\right)}\alpha_{1}^{\star}\alpha_{2}\alpha_{3}+g_{1}^{2}g_{2}^{\star2}mn\left|\alpha_{2}\right|^{2\left(n-2\right)}\left|\alpha_{3}\right|^{2\left(m-2\right)}\right.\\
 & \times & \alpha_{1}^{\star2}\alpha_{2}^{2}\alpha_{3}^{2}\left\{ \frac{1}{2}\left(m-1\right)\left(n-1\right)+\left(m-1\right)\left|\alpha_{2}\right|^{2}+\left(n-1\right)\left|\alpha_{3}\right|^{2}\right\} +h_{2}h_{3}^{\star}m^{2}n\alpha_{1}^{2}\alpha_{2}^{\star}\alpha_{4}^{\star}\\
 & \times & \left|\alpha_{2}\right|^{2\left(n-1\right)}\left|\alpha_{3}\right|^{2\left(m-1\right)}+g_{1}g_{4}^{\star}mn\left|\alpha_{2}\right|^{2\left(n-1\right)}\left|\alpha_{3}\right|^{2\left(m-2\right)}\alpha_{2}\alpha_{3}^{2}\alpha_{4}^{\star}\left(2\left|\alpha_{3}\right|^{2}+m-1\right)\\
 & + & \left.h_{1}^{\star2}h_{2}h_{3}mn\left(m-1\right)\left|\alpha_{1}\right|^{2}\left|\alpha_{2}\right|^{2\left(n-1\right)}\left|\alpha_{3}\right|^{2\left(m-2\right)}\alpha_{2}^{\star}\alpha_{3}^{\star2}\alpha_{4}+{\rm c.c.}\right].
\end{array}\label{bc}
\end{equation}

\begin{equation}
\begin{array}{lcl}
\left(\begin{array}{c}
E_{a,c}^{n,m}\\
E_{a,c}^{\prime n,m}
\end{array}\right) & = & \left|f_{2}\right|^{2}m\left|\alpha_{1}\right|^{2n}\left|\alpha_{3}\right|^{2\left(m-1\right)}\left(m\left|\alpha_{1}\right|^{2}\mp n\left|\alpha_{3}\right|^{2}\right)\\
 & + & \left|f_{3}\right|^{2}\left|\alpha_{1}\right|^{2\left(n-1\right)}\left|\alpha_{3}\right|^{2\left(m-1\right)}\left\{ m^{2}\left(1\pm2n\right)\left|\alpha_{1}\right|^{2}\left|\alpha_{4}\right|^{2}+n^{2}\left(1\pm2m\right)\left|\alpha_{3}\right|^{2}\left|\alpha_{4}\right|^{2}\right.\\
 & \mp & \left.mn\left|\alpha_{1}\right|^{2}\left|\alpha_{3}\right|^{2}+m^{2}n^{2}\left|\alpha_{4}\right|^{2}\right\} \pm\left[f_{1}f_{3}^{\star}mn\left|\alpha_{1}\right|^{2\left(n-1\right)}\alpha_{1}\left|\alpha_{3}\right|^{2\left(m-1\right)}\alpha_{3}\alpha_{4}^{\star}\right.\\
 & + & h_{2}^{\star}h_{3}m^{2}n\alpha_{1}^{\star2}\left|\alpha_{1}\right|^{2\left(n-1\right)}\alpha_{2}\left|\alpha_{3}\right|^{2\left(m-1\right)}\alpha_{4}+f_{2}f_{3}^{\star}mn^{2}\left|\alpha_{1}\right|^{2\left(n-1\right)}\alpha_{2}\left|\alpha_{3}\right|^{2\left(m-1\right)}\alpha_{3}^{2}\alpha_{4}^{\star}\\
 & + & f_{1}^{\star2}f_{3}^{2}mn\alpha_{1}^{\star2}\left|\alpha_{1}\right|^{2\left(n-2\right)}\alpha_{3}^{\star2}\left|\alpha_{3}\right|^{2\left(m-2\right)}\alpha_{4}^{2}\left\{ \left(n-1\right)\left|\alpha_{3}\right|^{2}+\left(m-1\right)\left|\alpha_{1}\right|^{2}+\frac{\left(m-1\right)\left(n-1\right)}{2}\right\} \\
 & + & f_{1}^{\star}f_{2}h_{1}^{\star}h_{3}mn\left(n-1\right)\alpha_{1}^{\star2}\left|\alpha_{1}\right|^{2\left(n-2\right)}\alpha_{2}\left|\alpha_{3}\right|^{2m}\alpha_{4}\\
 & + & \left.f_{1}f_{3}^{\star}h_{1}h_{2}^{\star}mn\left(m-1\right)\left|\alpha_{1}\right|^{2n}\alpha_{2}\left|\alpha_{3}\right|^{2\left(m-2\right)}\alpha_{3}^{2}\alpha_{4}^{\star}+c.c.\right]
\end{array}\label{ac}
\end{equation}

\begin{equation}
\begin{array}{lcl}
\left(\begin{array}{c}
E_{a,d}^{n,m}\\
E_{a,d}^{\prime n,m}
\end{array}\right) & = & \left|f_{3}\right|^{2}n\left|\alpha_{1}\right|^{2\left(n-1\right)}\left|\alpha_{4}\right|^{2m}\left(n\left|\alpha_{4}\right|^{2}\mp m\left|\alpha_{1}\right|^{2}\right)\end{array}\label{ad}
\end{equation}

\begin{equation}
\begin{array}{lcl}
\left(\begin{array}{c}
E_{c,d}^{n,m}\\
E_{c,d}^{\prime n,m}
\end{array}\right) & = & \left|h_{2}\right|^{2}n^{2}\left|\alpha_{1}\right|^{2}\left|\alpha_{3}\right|^{2\left(n-1\right)}\left|\alpha_{4}\right|^{2m}+\left|l_{2}\right|^{2}\left|\alpha_{3}\right|^{2\left(n-1\right)}\left|\alpha_{4}\right|^{2m}\left[n^{2}\left|\alpha_{4}\right|^{2}\mp mn\left|\alpha_{3}\right|^{2}\right]\end{array}\label{cd}
\end{equation}

\begin{equation}
\begin{array}{lcl}
\left(\begin{array}{c}
E_{bd}^{n,m}\\
E_{b,d}^{\prime n,m}
\end{array}\right) & = & \left|g_{2}\right|^{2}n^{2}\left|\alpha_{1}\right|^{2}\left|\alpha_{2}\right|^{2\left(n-1\right)}\left|\alpha_{4}\right|^{2m}\pm\left[l_{1}^{\star}l_{3}mn\alpha_{1}^{2}\alpha_{2}^{\star}\left|\alpha_{2}\right|^{2\left(n-1\right)}\alpha_{4}^{\star}\left|\alpha_{4}\right|^{2\left(m-1\right)}+c.c.\right]\end{array}\label{bd}
\end{equation}

\end{widetext} Here we would like to note that once we obtain analytic
expressions for $E_{i,j}^{n,m}$ and $E_{i,j}^{\prime n,m}$ in stimulated
Raman process, it is straightforward to study the special cases: (i)
spontaneous Raman process, where $\alpha_{2}=\alpha_{3}=\alpha_{4}=0,$
but $\alpha_{1}\neq0,$ and (ii) partially spontaneous Raman process,
where $\alpha_{1}\neq0$ and any one/two of the other three $\alpha_{i}\,(i=2,\,3,\,4)$
is/are non-zero. It is clear from the Eqs. (\ref{ab}-\ref{bd}) that
for spontaneous Raman process Eqs. (\ref{ab}-\ref{bd}) reduces to
zero. Hence for the spontaneous Raman process, no signature of intermodal
entanglement is observed. To investigate the possibility of higher
order intermodal entanglement in the stimulated Raman process we have
used $\chi=g=10^{4}$ Hz,
$|\alpha_{1}|=10,$ $|\alpha_{2}|=8,$ $|\alpha_{3}|=0.01,$ $|\alpha_{4}|=1$
\cite{foot2}. We have plotted the right hand side of (\ref{ab} -\ref{bd})
in Fig. \ref{fig:HZ2-1} and Fig. \ref{fig:HZ2}
for $m=1$and $n=1,\,2$ and $3.$We observed that HZ-1 criteria can
detect the higher order intermodal entanglement in the stimulated
Raman process for different values of the phase angle or all phase
angles of the input pump field (i.e., for $\phi=0,\,\frac{\pi}{2}$
and $\pi$) for all the possible modes except pump-phonon ($ac$)
and phonon-anti-Stokes ($cd$ ) modes. It is interesting to note that
higher order intermodal entanglement is observed in pump-Stokes mode,
although in the lowest order it was not observed. Further, the figures
show that the depths of the nonclassicality parameters $E_{i,j}^{n,m}$
and $E_{i,j}^{\prime n,m}$ increase with the order. Use of HZ-1 criteria
also led to similar features in the partially spontaneous Raman process
(not in figure). In other words, we observed signatures of intermodal
entanglement in all the cases except pump-phonon ($ac$) and phonon-anti-Stokes
($cd$) modes. As HZ-1 is only a sufficient (not necessary) criterion,
it may have failed to witness entanglement, keeping this fact in mind,
we have plotted the right hand side of Eq. (\ref{ab}- \ref{bd})
using HZ-2 criteria (See Fig. \ref{fig:HZ2}). It is interesting to
note that HZ-2 criterion can detect the higher order intermodal entanglement
in pump-phonon ($ac$) mode for phase angle $\phi=\frac{\pi}{2}$,
which was not detected by HZ-1 criterion, in the stimulated and partial
spontaneous Raman processes. However, we do not observe any signature
of higher order intermodal entanglement for spontaneous Raman process.
Thus, the stimulated Raman process provides a very nice example of
a physical system which can produce higher order entanglement. 

\subsection{Three mode entanglement}

\begin{figure}[h]
\includegraphics[scale=0.4]{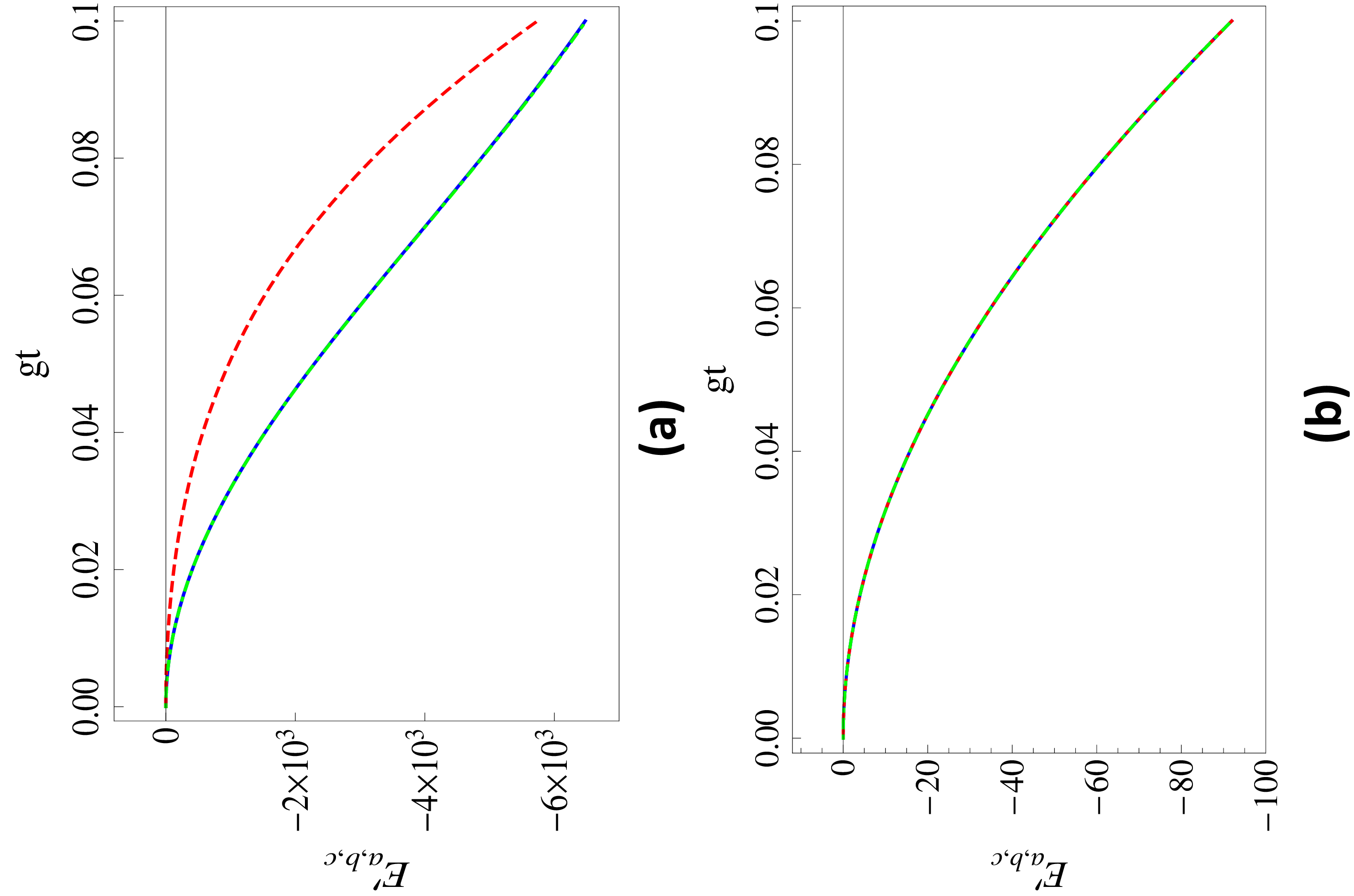}

\caption{\label{fig:threemode}(Colour online) The variation of three mode
entanglement among pump, Stokes and vibration phonon modes (a) for
stimulated Raman process using $\left|\alpha_{1}\right|=10,$ $\left|\alpha_{2}\right|=8,$
$\left|\alpha_{3}\right|=0.01$, $\left|\alpha_{4}\right|=1$ (b)
for spontaneous Raman process using $\left|\alpha_{1}\right|=10,$
$\left|\alpha_{2}\right|=\left|\alpha_{3}\right|=\left|\alpha_{4}\right|=0$
with the smooth, dashed and dash-dotted lines corresponding to $\phi=0,$
$\frac{\pi}{2}$ and $\pi,$ respectively.}
\end{figure}

There exists another alternative way to study the higher-order entanglement.
To be precise, all multi-mode entanglements are essentially higher-order
entanglement. In other words, three mode entanglement always indicates
higher order entanglement. In order to investigate the three mode
entanglement, we use the following criterion \cite{Ent condition-multimode}
\begin{equation}
E_{a,b,c}^{\prime}=\begin{array}{lcl}
\begin{array}{lcl}
\left\langle N_{a}\right\rangle \left\langle N_{b}\right\rangle \left\langle N_{c}\right\rangle  & - & \left|\left\langle abc\right\rangle \right|^{2}\end{array} & < & 0\end{array},\label{eq:threemodecriteria}
\end{equation}
where $\left\langle N_{a}\right\rangle ,$ $\left\langle N_{b}\right\rangle ,$
and $\left\langle N_{c}\right\rangle $ are average value of the number
operators of the pump mode Stokes mode and vibration phonon mode respectively.
Using equations (\ref{soln1}), (\ref{eq:initial state}) and (\ref{eq:threemodecriteria})
we obtain\begin{widetext} 
\begin{equation}
\begin{array}{lcl}
E_{a,b,c}^{\prime} & = & \left\langle N_{a}\right\rangle \left\langle N_{b}\right\rangle \left\langle N_{c}\right\rangle -\left|\left\langle abc\right\rangle \right|^{2}\\
 & = & \left|f_{2}\right|^{2}\left(5\left|\alpha_{1}\right|^{2}\left|\alpha_{2}\right|^{2}\left|\alpha_{3}\right|^{2}-\left|\alpha_{1}\right|^{4}\left|\alpha_{3}\right|^{2}-\left|\alpha_{1}\right|^{4}-\left|\alpha_{1}\right|^{4}\left|\alpha_{2}\right|^{2}\right)\\
 & + & \left|f_{3}\right|^{2}\left(\left|\alpha_{1}\right|^{2}\left|\alpha_{2}\right|^{2}\left|\alpha_{3}\right|^{2}-4\left|\alpha_{2}\right|^{2}\left|\alpha_{4}\right|^{2}-3\left|\alpha_{2}\right|^{2}\left|\alpha_{3}\right|^{2}\left|\alpha_{4}\right|^{2}-3\left|\alpha_{1}\right|^{2}\left|\alpha_{2}\right|^{2}\left|\alpha_{4}\right|^{2}\right)+\\
 & - & \left[h_{1}h_{2}^{\star}\alpha_{1}^{\star}\left|\alpha_{1}\right|^{2}\alpha_{2}\alpha_{3}+2f_{1}f_{3}^{\star}\alpha_{1}\left|\alpha_{2}\right|^{2}\alpha_{3}\alpha_{4}^{\star}+h_{2}h_{3}^{\star}\left(2\alpha_{1}^{2}\alpha_{2}^{\star}\alpha_{4}^{\star}\right.\right.\\
 & + & \left.\left.\left|\alpha_{1}\right|^{2}\alpha_{1}^{2}\alpha_{2}^{\star}\alpha_{4}^{\star}+2\alpha_{1}^{2}\alpha_{2}^{\star}\left|\alpha_{2}\right|^{2}\alpha_{4}^{\star}\right)+{\rm c.c.}\right].
\end{array}\label{eq:threemode}
\end{equation}
For the spontaneous Raman process, Eq. (\ref{eq:threemode}) reduces
to 
\begin{equation}
\begin{array}{lcl}
E_{a,b,c}^{\prime} & = & -\left|f_{2}\right|^{2}\left|\alpha_{1}\right|^{4}\end{array},\label{threemodespon}
\end{equation}
which is clearly negative and thus indicate the existence of tripartite
entanglement in the spontaneous Raman process.

\end{widetext} To investigate the existence of three mode entanglement
in the stimulated Raman process, we plot the right hand side of the
equation (\ref{eq:threemode}) in Fig.\ref{fig:threemode} for three
different values of the phase angle of the input pump field, i.e.,
for $\phi=0$ (blue smooth line), $\phi=\frac{\pi}{2}$ (red dotted
line) and $\phi=\pi$ (green dash dotted line). The negative regions
of the plots clearly illustrate the existence of tri-modal (higher
order) entanglement. From Fig.\ref{fig:threemode} 
we can clearly observe the signature of higher order entanglement
for different values of phase angle of the input pump field for stimulated
and spontaneous Raman processes.

\subsection{Four mode entanglement}

In order to investigate the four mode entanglement we use the following
criterion, which is similar to that of Li et al.'s three mode criterion
\cite{Ent condition-multimode}: 
\begin{equation}
\begin{array}{lcl}
E_{a,b,c,d}^{\prime} & = & \left\langle N_{a}\right\rangle \left\langle N_{b}\right\rangle \left\langle N_{c}\right\rangle \left\langle N_{d}\right\rangle -\left|\left\langle abcd\right\rangle \right|^{2}<0,\end{array}
\end{equation}
where $a,$$b,$$c$ and $d$ are arbitrary operators and the negative
value of $E_{a,b,c,d}^{\prime}$ gives the signature of the higher
order entanglement. Now, we investigate the higher order entanglement
i.e., the entanglement among the four modes of the stimulated Raman
and spontaneous Raman processes and we obtain \begin{widetext} 
\begin{equation}
\begin{array}{lcl}
E_{a,b,c,d}^{\prime} & = & \left|f_{2}\right|^{2}\left|\alpha_{1}\right|^{2}\left(5\left|\alpha_{2}\right|^{2}\left|\alpha_{3}\right|^{2}\left|\alpha_{4}\right|^{2}-\left|\alpha_{1}\right|^{2}\left|\alpha_{3}\right|^{2}\left|\alpha_{4}\right|^{2}-\left|\alpha_{1}\right|^{2}\left|\alpha_{4}\right|^{2}\right)\\
 & + & \left|f_{3}\right|^{2}\left|\alpha_{2}\right|^{2}\left|\alpha_{4}\right|^{2}\left(7\left|\alpha_{1}\right|^{2}\left|\alpha_{3}\right|^{2}-4\left|\alpha_{4}\right|^{2}-3\left|\alpha_{1}\right|^{2}\left|\alpha_{4}\right|^{2}-3\left|\alpha_{3}\right|^{2}\left|\alpha_{4}\right|^{2}\right)\\
 & - & \left[h_{1}^{\star}h_{2}\left|\alpha_{1}\right|^{2}\alpha_{1}\alpha_{2}^{\star}\alpha_{3}^{\star}\left|\alpha_{4}\right|^{2}+2f_{1}^{\star}f_{3}\alpha_{1}^{\star}\left|\alpha_{2}\right|^{2}\alpha_{3}^{\star}\left|\alpha_{4}\right|^{2}\alpha_{4}+f_{2}f_{3}^{\star}\left|\alpha_{2}\right|^{2}\alpha_{2}\alpha_{3}^{2}\alpha_{4}^{\star}\left|\alpha_{4}\right|^{2}\right.\\
 & + & \left(l_{1}^{\star}l_{3}-h_{2}h_{3}^{\star}\right)\left|\alpha_{1}\right|^{2}\alpha_{1}^{2}\alpha_{2}^{\star}\left|\alpha_{3}\right|^{2}\alpha_{4}^{\star}+h_{2}h_{3}^{\star}\left(2\alpha_{1}^{2}\alpha_{2}^{\star}\alpha_{4}^{\star}\left|\alpha_{4}\right|^{2}+\left|\alpha_{1}\right|^{2}\alpha_{1}^{2}\alpha_{2}^{\star}\alpha_{4}^{\star}\left|\alpha_{4}\right|^{2}\right.\\
 & + & \left.2\alpha_{1}^{2}\alpha_{2}^{\star}\left|\alpha_{2}\right|^{2}\alpha_{4}^{\star}\left|\alpha_{4}\right|^{2}+3\alpha_{1}^{2}\alpha_{2}^{\star}\left|\alpha_{3}\right|^{2}\alpha_{4}^{\star}\left|\alpha_{4}\right|^{2}\right)+f_{1}f_{2}^{\star}h_{1}^{\star}h_{2}\alpha_{1}^{2}\alpha_{2}^{\star2}\alpha_{3}^{\star2}\left|\alpha_{4}\right|^{2}\\
 & + & h_{1}^{\star}h_{2}l_{1}l_{2}^{\star}\left(\left|\alpha_{1}\right|^{4}\alpha_{2}^{\star}\alpha_{3}^{\star2}\alpha_{4}+\alpha_{2}^{\star}\left|\alpha_{2}\right|^{2}\alpha_{3}^{\star2}\left|\alpha_{4}\right|^{2}\alpha_{4}\right)+2f_{1}^{\star}f_{3}l_{1}l_{2}^{\star}\alpha_{1}^{\star2}\left|\alpha_{2}\right|^{2}\alpha_{3}^{\star2}\alpha_{4}^{2}\\
 & + & \left.\left(h_{1}^{\star}h_{4}+3g_{1}^{\star}g_{2}h_{1}^{\star}h_{3}\right)\left|\alpha_{1}\right|^{2}\alpha_{2}^{\star}\alpha_{3}^{\star2}\left|\alpha_{4}\right|^{2}\alpha_{4}+c.c.\right]
\end{array}\label{entaglemntabcd}
\end{equation}

\end{widetext}

\begin{figure}[h]
\includegraphics[scale=0.4]{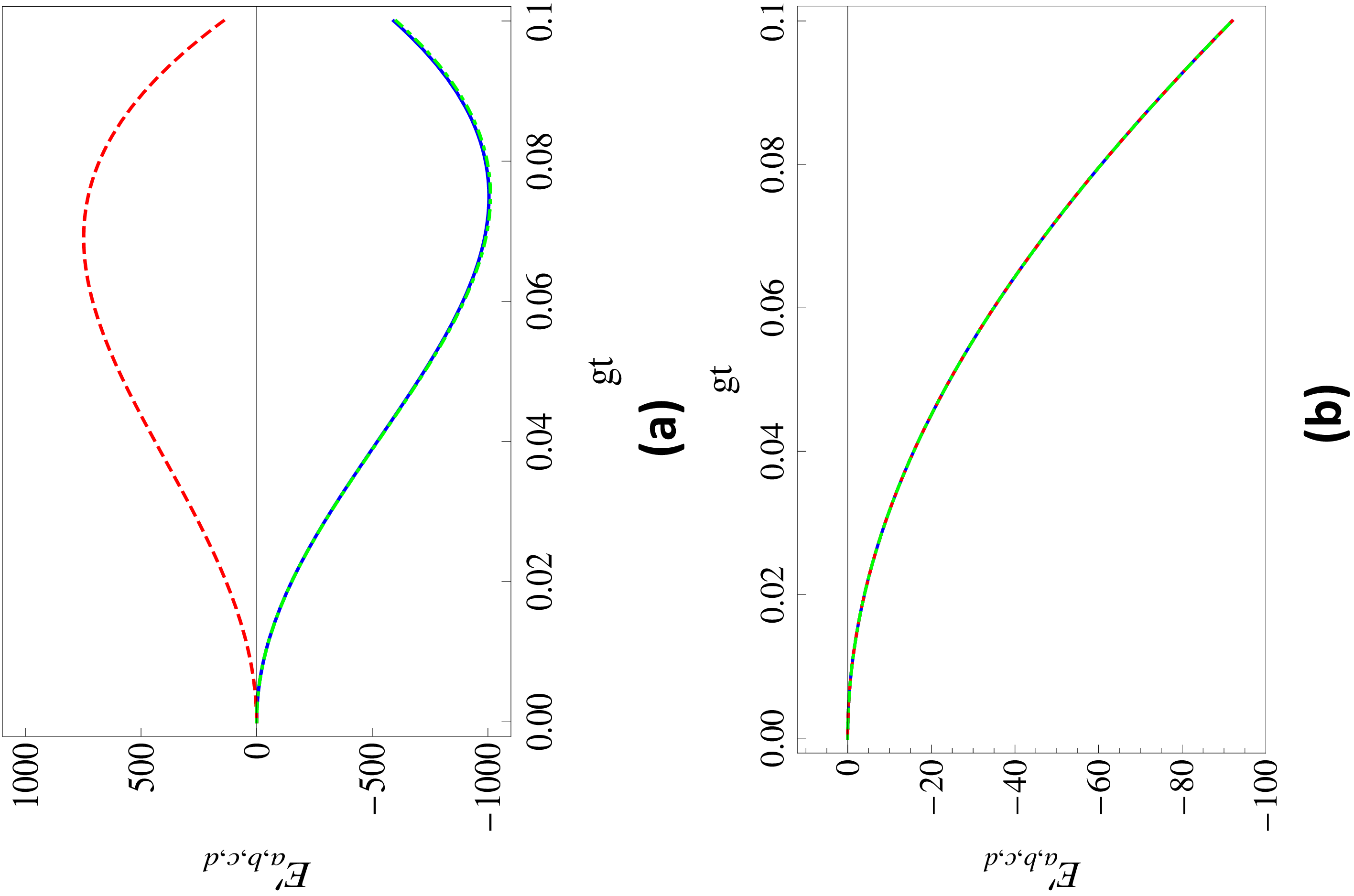}

\caption{\label{fig:fourmode}(Colour online) Four mode entanglement among
pump, Stokes, vibration phonon and anti-Stokes modes is depicted in
(a) stimulated Raman process using $\left|\alpha_{1}\right|=10,$
$\left|\alpha_{2}\right|=8,$ $\left|\alpha_{3}\right|=0.01$, $\left|\alpha_{4}\right|=1$
(b) partial spontaneous Raman process using $\left|\alpha_{1}\right|=10,$
$\left|\alpha_{2}\right|=\left|\alpha_{3}\right|=0$. Here, the smooth
line, dashed line and dash-dotted line corresponds to $\phi=0,$ $\frac{\pi}{2}$
and $\pi,$ respectively. }
\end{figure}

In order to investigate the possibility of observing 4-mode entanglement
in the Raman processes, in Fig. \ref{fig:fourmode}, we have plotted
the variation of right hand side of Eq. (\ref{entaglemntabcd}) with
the rescaled time $gt$. Quite interestingly, for appropriate choice
of the phase of the pump mode, 4 mode entanglement is observed in
both stimulated Raman process and partially spontaneous Raman process.

\section{Conclusions\label{sec:Conclusions}}

Recently, nonclassical properties of the stimulated Raman process
have been extensively studied by some of the present authors \cite{Anirban with Perina,raman-pra}.
In those studies intermodal entanglement in different modes of the
stimulated Raman process was reported. Intermodal entanglement between
Stokes mode and the vibration mode in the Raman processes was also
reported by Kuznetsov \cite{kuznetsov}. However, higher order entanglement
was not investigated. In the present paper higher order entanglement
in stimulated Raman process is studied in detail and the observed
higher order entanglement are illustrated through the negative regions
of the Figs.\ref{fig:HZ2-1}-\ref{fig:fourmode}. In Figs. \ref{fig:HZ2-1}
and \ref{fig:HZ2}, the existence of higher order two-mode entanglement
between various possible combinations of modes are illustrated using
HZ-1 criterion and HZ-2 criterion, respectively. Specifically, using
HZ-1 criterion, we have observed the intermodal higher order entanglement
for all the possible combinations of modes, except pump-phonon $(ac)$
and phonon-anti-Stokes $(cd)$ modes in stimulated Raman process (cf.
Fig.\ref{fig:HZ2-1}) and in partially spontaneous Raman processes
(not shown in figure). However, we found that HZ-2 criteria can detect
the signature of higher order intermodal entanglement only in Stokes-phonon
($bc$), pump-phonon ($ac$) and Stokes-anti-Stokes ($bd$) modes
in the stimulated and partially spontaneous Raman process Fig.\ref{fig:HZ2},
but it is interesting to note that HZ-2 criteria can detect the higher
order intermodal entanglement in pump-phonon ($ac$) mode whereas
HZ-1 criteria fails to detect this. Thus, by combining the results,
we have observed the existence of two-mode higher order entanglement
in stimulated and partially spontaneous Raman possesses in all possible
cases except in phonon-anti-Stokes $(cd)$ modes. However, no signature
of intermodal entanglement is observed for the spontaneous Raman process.
Another interesting point is that the present investigation reveals
the signature of higher order intermodal entanglement in pump-Stokes
mode ($ab$) in stimulated Raman process, but intermodal entanglement
in $ab$ modes were not observed in lowest order (cf. Fig. 2a, 3a
and 4a of Ref. \cite{raman-pra}). As all the multi-partite (multi-mode)
entanglement are essentially higher order entanglement, we investigated
the possibility of observing three mode and four mode entanglements
in Raman processes and found that tri-modal entanglement can be observed
among pump, Stokes and vibration phonon mode ($abc$) in both stimulated
and spontaneous Raman processes (cf. Fig.\ref{fig:threemode}), and
it is also possible to observe entanglement among four modes (pump,
Stokes, vibration phonon and anti-Stokes) in stimulated and partially
spontaneous Raman processes (see Fig.\ref{fig:fourmode}). As recently
many applications of multi-partite entanglement has been proposed,
we hope that the present observation on the possibility of observing
multi-mode entanglement in Raman process would be of help in realizing
some of the recently proposed schemes that are based on multi-partite
entanglement. Further, it is easy to experimentally realize Raman
process and thus the results reported here can be experimentally verified
using the available technologies.

Bosonic Hamiltonians similar to the one studied here frequently appear
in quantum optical, opto-mechanical and atomic systems. Thus, the
methodology adopted here may also be used in those systems to study
the existence of nonclassical states in normal
and higher order entanglement in particular. Keeping this in mind,
we conclude the present work with an expectation that this work would
lead to a bunch of similar studies in other bosonic systems. 
\begin{acknowledgments}
SKG acknowledges the financial support by the UGC,
Government of India in the framework of the UGC minor project no.
PSW-148/14-15 (ERO). AP and BS thanks Kishore Thapliyal for his feedback
on the work and his help in preparing the final figures.
\end{acknowledgments}
\appendix

\section{\label{appendix}Parameters for the solutions in Eq. (3)}

\begin{eqnarray}
f_{1} & = & \exp(-i\omega_{a}t),\notag\\
f_{2} & = & \frac{ge^{-i\omega_{a}t}}{\Delta\omega_{1}}\left[e^{-i\Delta\omega_{1}t}-1\right],\notag\\
f_{3} & = & -\frac{\chi e^{-i\omega_{a}t}}{\Delta\omega_{2}}\left[e^{i\Delta\omega_{2}t}-1\right],\notag\\
f_{4} & = & \begin{array}{l}
-\frac{\chi ge^{-i\omega_{a}t}}{\Delta\omega_{1}}\left[\frac{e^{-i(\Delta\omega_{1}-\Delta\omega_{2})t}-1}{\Delta\omega_{1}-\Delta\omega_{2}}+\frac{e^{i\Delta\omega_{2}t}}{\Delta\omega_{2}}\right]\\
-\frac{\chi ge^{-i\omega_{a}t}}{\Delta\omega_{2}}\left[\frac{e^{-i(\Delta\omega_{1}-\Delta\omega_{2})t}-1}{\Delta\omega_{1}-\Delta\omega_{2}}-\frac{e^{-i\Delta\omega_{1}t}}{\Delta\omega_{1}}\right],
\end{array}\label{f}\\
f_{5} & = & \frac{g^{2}e^{-i\omega_{a}t}}{\Delta\omega_{1}^{2}}\left[e^{-i\Delta\omega_{1}t}-1\right]+\frac{ig^{2}te^{-i\omega_{a}t}}{\Delta\omega_{1}},\notag\\
f_{6} & = & f_{5},\notag\\
f_{7} & = & \frac{\chi^{2}e^{-i\omega_{a}t}}{\Delta\omega_{2}^{2}}\left[e^{i\Delta\omega_{2}t}-1\right]-\frac{i\chi^{2}te^{-i\omega_{a}t}}{\Delta\omega_{2}},\notag\\
f_{8} & = & -f_{7}.\notag
\end{eqnarray}

\begin{eqnarray}
g_{1} & = & \exp(-i\omega_{b}t),\notag\\
g_{2} & = & -\frac{ge^{-i\omega_{b}t}}{\Delta\omega_{1}}\left[e^{i\Delta\omega_{1}t}-1\right],\notag\\
g_{3} & = & \begin{array}{l}
\frac{\chi ge^{-i\omega_{b}t}}{\Delta\omega_{2}(\Delta\omega_{1}-\Delta\omega_{2})}\left[e^{i(\Delta\omega_{1}-\Delta\omega_{2})t}-1\right]\\
-\frac{\chi ge^{-i\omega_{b}t}}{\Delta\omega_{2}\Delta\omega_{1}}\left[e^{i\Delta\omega_{1}t}-1\right],
\end{array}\label{g}\\
g_{4} & = & \begin{array}{l}
\frac{\chi ge^{-i\omega_{b}t}}{\Delta\omega_{2}(\Delta\omega_{1}+\Delta\omega_{2})}\left[e^{i(\Delta\omega_{1}+\Delta\omega_{2})t}-1\right]\\
-\frac{\chi ge^{-i\omega_{b}t}}{\Delta\omega_{2}\Delta\omega_{1}}\left[e^{i\Delta\omega_{1}t}-1\right],
\end{array}\notag\\
g_{5} & = & \frac{g^{2}e^{-i\omega_{b}t}}{\Delta\omega_{1}^{2}}\left[e^{i\Delta\omega_{1}t}-1\right]-\frac{ig^{2}te^{-i\omega_{b}t}}{\Delta\omega_{1}},\notag\\
g_{6} & = & -g_{5}.\notag
\end{eqnarray}

\begin{eqnarray}
h_{1} & = & \exp(-i\omega_{c}t)\notag\\
h_{2} & = & -\frac{ge^{-i\omega_{c}t}}{\Delta\omega_{1}}\left[e^{i\Delta\omega_{1}t}-1\right]\notag\\
h_{3} & = & -\frac{\chi e^{-i\omega_{c}t}}{\Delta\omega_{2}}\left[e^{i\Delta\omega_{2}t}-1\right]\notag\\
h_{4} & = & \begin{array}{l}
\frac{\chi ge^{-i\omega_{c}t}}{\Delta\omega_{2}}\left[\frac{e^{i(\Delta\omega_{1}+\Delta\omega_{2})t}-1}{\Delta\omega_{1}+\Delta\omega_{2}}-\frac{e^{i\Delta\omega_{1}t}}{\Delta\omega_{1}}\right]\\
-\frac{\chi ge^{-i\omega_{c}t}}{\Delta\omega_{1}}\left[\frac{e^{i(\Delta\omega_{1}+\Delta\omega_{2})t}-1}{\Delta\omega_{1}+\Delta\omega_{2}}-\frac{e^{i\Delta\omega_{2}t}}{\Delta\omega_{2}}\right]
\end{array}\label{h}\\
h_{5} & = & -\frac{g^{2}e^{-i\omega_{c}t}}{\Delta\omega_{1}^{2}}\left[e^{i\Delta\omega_{1}t}-1\right]+\frac{ig^{2}te^{-i\omega_{c}t}}{\Delta\omega_{1}}\notag\\
h_{6} & = & -h_{5}\notag\\
h_{7} & = & -\frac{\chi^{2}e^{-i\omega_{c}t}}{\Delta\omega_{2}^{2}}\left[e^{i\Delta\omega_{2}t}-1\right]+\frac{i\chi^{2}te^{-i\omega_{c}t}}{\Delta\omega_{2}}\notag\\
h_{8} & = & \frac{\chi^{2}e^{-i\omega_{c}t}}{\Delta\omega_{2}^{2}}\left[e^{i\Delta\omega_{2}t}-1\right]-\frac{i\chi^{2}te^{-i\omega_{c}t}}{\Delta\omega_{2}}\notag
\end{eqnarray}

\begin{eqnarray}
l_{1} & = & \exp(-i\omega_{d}t)\notag\\
l_{2} & = & \frac{\chi e^{-i\omega_{d}t}}{\Delta\omega_{2}}\left[e^{-i\Delta\omega_{2}t}-1\right]\notag\\
l_{3} & = & \begin{array}{l}
\frac{\chi ge^{-i\omega_{d}t}}{\Delta\omega_{1}(\Delta\omega_{1}-\Delta\omega_{2})}\left[e^{i(\Delta\omega_{1}-\Delta\omega_{2})t}-1\right]\\
+\frac{\chi ge^{-i\omega_{d}t}}{\Delta\omega_{2}\Delta\omega_{1}}\left[e^{-i\Delta\omega_{2}t}-1\right]
\end{array}\label{l}\\
l_{4} & = & \begin{array}{l}
\frac{\chi ge^{-i\omega_{d}t}}{\Delta\omega_{1}(\Delta\omega_{1}+\Delta\omega_{2})}\left[e^{-i(\Delta\omega_{1}+\Delta\omega_{2})t}-1\right]\\
-\frac{\chi ge^{-i\omega_{d}t}}{\Delta\omega_{2}\Delta\omega_{1}}\left[e^{-i\Delta\omega_{2}t}-1\right]\,
\end{array}\notag\\
l_{5} & = & \frac{i\chi^{2}te^{-i\omega_{d}t}}{\Delta\omega_{2}}+\frac{\chi^{2}e^{-i\omega_{d}t}}{\Delta\omega_{2}^{2}}\left[e^{-i\Delta\omega_{2}t}-1\right]\notag\\
l_{6} & = & l_{5}\notag
\end{eqnarray}


\begin{thebibliography}{10}
\bibitem{foot2} Same values of $\chi,$ $g$ and $|\alpha_{i}|$
are used in the entire paper (unless otherwise specified). For spontaneous
and partially spontaneous Raman proceses these values of $|\alpha_{i}|$
are used for non-zero $|\alpha_{i}|$'s.
\end{thebibliography}
\end{document}